\documentclass[a4paper,12pt]{article}

\usepackage{graphicx,amssymb,bm,latexsym}

\makeatletter
\@addtoreset{equation}{section}

\makeatother
\newcommand{\bea}{\begin{eqnarray}}
\newcommand{\ena}{\end{eqnarray}}
\newcommand{\vs}[1]{\vspace{#1 mm}}
\newcommand{\hs}[1]{\hspace{#1 mm}}
\renewcommand{\a}{\alpha}
\renewcommand{\b}{\beta}
\renewcommand{\c}{\gamma}
\renewcommand{\d}{\delta}
\newcommand{\e}{\epsilon}

\newcommand{\dsl}{\pa \kern-0.5em /}

\newcommand{\shalf}{\frac{1}{2}}
\newcommand{\pa}{\partial}
\newcommand{\nn}{\nonumber\\}
\newcommand{\p}[1]{(\ref{#1})}

\begin{document}
\renewcommand{\thefootnote}{\fnsymbol{footnote}}

\begin{titlepage}
\setcounter{page}{0}
\begin{flushright}
OU-HET 448 \\
IP/BBSR/2003-14\\
hep-th/0306186
\end{flushright}

\vs{15}
\begin{center}
{\Large\bf Intersecting branes in pp-wave spacetime}

\vs{15}
{\large
Nobuyoshi Ohta$^a$\footnote{e-mail: ohta@phys.sci.osaka-u.ac.jp},
Kamal L. Panigrahi$^b$\footnote{e-mail: kamal@iopb.res.in} and
Sanjay$^b$\footnote{email: sanjay@iopb.res.in, sanjay@imsc.res.in}\footnote{Now at Institute of Mathematical Sciences, Chennai-600 113, India}}\\
\vs{10}

{\em $^a$Department of Physics, Osaka University,
Toyonaka, Osaka 560-0043, Japan\\
\vs{5}

$^{b}$Institute of Physics, Sachivalya Marg, Bhubaneswar 751 005, India}
\end{center}
\vs{15}
\centerline{{\bf{Abstract}}}
\vs{5}

We derive intersecting brane solutions in pp-wave spacetime by solving
the supergravity field equations explicitly. The general intersection rules
are presented. We also generalize the construction to the non-extremal
solutions. Both the extremal and non-extremal solutions presented here
are asymptotic to BFHP plane waves. We find that these solutions
do not have regular horizons.
\end{titlepage}

\newpage
\renewcommand{\thefootnote}{\arabic{footnote}}
\setcounter{footnote}{0}

\setcounter{page}{2}
\section{Introduction}

PP-wave spacetime provides an example of exact string theory background.
It has been known for some time that pp-wave spacetime yields exact classical
backgrounds for string theory, with all $\alpha'$ corrections vanishing
\cite{klim,steif}. Also it has been shown recently that these backgrounds
are exactly solvable in the light cone gauge~\cite{met0}-\cite{met2}.
Many of these backgrounds are obtained in the Penrose limit~\cite{penrose}
of $AdS_p\times S^q$ type of geometry and preserve at least the same amount
of supersymmetry as that of parent background~\cite{bfhp,blau}.
PP-wave backgrounds have also proven an interesting place to test the ideas
of holography. It was conjectured in \cite{maldacena} that the type IIB
string theory in $AdS_5\times S^5$ background is dual to the ${\cal N}=4$
super Yang-Mills theory on the boundary of $AdS_5$.
Partly motivated by this, Berenstein, Maldacena and Nastase (BMN)
\cite{malda} have argued that a particular sector of ${\cal N} = 4$
super Yang-Mills theory containing  the operators with large $R$
charge $J$, is dual to Type IIB string theory on pp-wave background
with RR flux. The fact that the string theory in pp-wave background
is exactly solvable has opened up the window to understand the duality
beyond the supergravity limit. The correspondence between the string
states and black holes is also adapted to pp-wave backgrounds~\cite{li}.

D-branes can probe the nonperturbative dynamics of the string theory and
they have been used to study various duality aspects of string theory.
Several aspects of D-branes in pp-wave spacetime e.g. the supergravity
solutions, open string spectrum and the supersymmetric properties of
branes and their bound states are studied extensively in the
past~\cite{dabh}-\cite{tay}. Recently the focus has been on the most
general pp-wave background with non-constant flux turned on
\cite{maoz}-\cite{bonelli}.
D-brane solutions in these backgrounds and their supersymmetry
properties are studied in refs.~\cite{hikida,kamal1}. In view of these
recent developments in the study of D-branes and thermodynamics of strings
in pp-wave background, it is interesting to give a systematic derivation
of the general D-brane solutions in the pp-wave backgrounds.

In this paper we present a general class of intersecting brane solutions
in the pp-wave background by using the method developed in ref.~\cite{NO}.
We start with a general ansatz for the metric and solve for the field equations
of the supergravity. We also derive the intersection rules for the branes
in this background. This is the pp-wave generalization of intersection rules
for $p$-branes derived in \cite{NO} (see also \cite{papado}-\cite{aeh}
for discussion of brane intersections). The method used here also applies
to other brane solutions~\cite{NO1} and is quite useful.

The existence of black holes or so-called black branes in pp-wave
spacetime is also discussed recently. In particular there are
`no-go theorems' for the existence of horizons in pp-wave spacetimes
admitting a covariantly constant null isometry~\cite{rangamani}-\cite{liu}.
The covariant constancy condition is
further relaxed and a similar `no-go theorem' is proved in \cite{liu}
for spacetimes which are asymptotic to plane wave spacetime. Though, a
version of `no-go theorem' is still lacking for the backgrounds with
sources and admitting null isometry, some examples are studied in
ref.~\cite{liu}. It has been found
that in pp-wave spacetime supported by non-zero 5-form flux, while
some of the extremal solutions admit horizons, the corresponding
non-extremal deformations result in naked singularity.\footnote{However,
if we further relax the null isometry condition, it might be possible to
find black branes with regular event horizon in pp-wave spacetime~\cite{liu}
(see also the recent paper by Gimon et. al. \cite{gimon}).}
We further examine the tidal force in the parallel transported frame
and find that these solutions all do not have regular horizons~\cite{bre}.

Here we look for the non-extremal deformations of brane solutions in
pp-wave spacetime supported by NS-NS 3-form flux. We are interested in
pp-wave background of the asymptotic form
\bea
ds_D^2 & = & -2 dudv + K(y_\a, z_i) du^2 + \sum_{\a=2}^{d-1}dy_\a^2
+ \sum_{i=1}^{\tilde d +2}dz_i^2, \nn
H^{(3)} & = &\pa_{j} b_{k}(z_i)~ du\wedge dz^{j}\wedge dz^{k},
\label{plane}
\ena
where $D=d+\tilde d+2$, the coordinates $u=(t+x)/\sqrt{2}$, $v=(t-x)/\sqrt{2}$
and $y_\a, (\a=2,\ldots, d-1)$ parameterize the $d$-dimensional world-volume
directions. The NS-NS 3-form, $H^{(3)}$ breaks the $SO(\tilde d+2)$ isometry
of transverse $z_i$ ($i=1,...,\tilde d +2$) directions.

The plan of the paper is as follows. In sect.~2, we present
classical solutions of intersecting branes in arbitrary dimension $D$
in pp-wave background in the presence of non-constant NS-NS 3-form flux,
switched on along the transverse directions to the branes. We also present
the intersection rules for branes in this background.
In sect.~3, we generalize the above construction to
non-extremal cases. The intersection rules along with
the `blackening' functions are derived by solving the Einstein equations.
Sect.~4 is devoted to the discussion on the possible
horizon and black hole solutions in the above pp-wave background.
We conclude in sect.~5 with some remarks.

\section{Extremal Solutions}

The low-energy effective action for the supergravity system coupled to
dilaton and $n_A$- form field strength is given by
\bea
I = \frac{1}{16 \pi G_D} \int d^D x \sqrt{-g} \left[
 R - \shalf (\pa \phi)^2 - \sum_{A=1}^m \frac{1}{2 n_A!} e^{a_A \phi}
 F_{n_A}^2 \right],
\label{action}
\ena
where $G_D$ is the Newton constant in $D$ dimensions and $g$ is the
determinant of the metric. The last term includes both RR and
NS-NS field strengths and $a_A = \shalf (5-n_A)$ for RR field
strengths and $a_A = -1$ for NS-NS 3-form. We put fermions and other
background fields to be zero.

{}From the action (\ref{action}), one can derive the field
equations/Bianchi identities
\bea
R_{\mu\nu} = \shalf \pa_\mu \phi \pa_\nu \phi + \sum_{A} \frac{1}{2 n_A!}
 e^{a_A \phi} \left[ n_A \left( F_{n_A}^2 \right)_{\mu\nu}
 - \frac{n_A -1}{D-2} F_{n_A}^2 g_{\mu\nu} \right],
\label{Einstein}
\ena
\bea
\Box \phi = \sum_{A} \frac{a_A}{2 n_A!} e^{a_A \phi} F_{n_A}^2,
\label{dila}
\ena
\bea
\pa_{\mu_1} \left( \sqrt{- g} e^{a_A \phi} F^{\mu_1 \cdots \mu_{n_A}} \right)
 = 0,
\label{field}
\ena
\bea
\pa _{[\mu} F_{\mu_1 \cdots \mu_{n_A}]} = 0.
\label{bianchi}
\ena

We start with the most general ansatz for the metric in pp-wave
spacetime consistent with the isometries of the background, and the
 NS-NS 3-form field strength:
\bea
ds_D^2 &=& e^{2u_0} [-2dudv + K(y_\a, z_i) du^2]
+  \sum_{\a=2}^{d-1} e^{2 u_\a} dy_\a^2
+e^{2B}[dr^2 + r^2 d\Omega_{\tilde d +1}^2],
\label{met}
\ena
where $D=d+\tilde d+2$, the coordinates $u$, $v$
and $y_\a, (\a=2,\ldots, d-1)$ parameterize the $d$-dimensional world-volume
directions and the remaining $\tilde d + 2 $ coordinates $z_i$ are
transverse to the brane world-volume; these are interchangeably
used here with the radial coordinate $r$ and $\tilde d+1$ angles.
$d\Omega_{\tilde d+1}^2$ is the line element of the
$(\tilde d+1)$-dimensional sphere. All the warp factors are assumed to
depend on $r$ only. In this and next sections, $d, \tilde d~ {\rm and}~D$
are general, but in sect.~4 we shall put $D=10$.

The general ansatz for the background for an electrically
charged $q_A$-brane is given by
\bea
F_{uv\a_2 \cdots \a_q r} = \e_{uv\a_2 \cdots \a_q} E', ~~~~~~~~~(n_A = q_A+2).
\label{elec}
\ena
Similarly, the magnetic case is given by
\bea
F^{\a_{q_A+1} \cdots \a_{d-1} a_1 \cdots a_{\tilde d+1}} = \frac{1}{\sqrt{-g}}
 e^{-a_A\phi} \e^{\a_{q_A + 1} \cdots \a_{d-1} a_1 \cdots a_{\tilde d+1}r}
{\tilde E}', \quad
 (n_A =D-q_A-2)
\label{mag}
\ena
where $a_1, \cdots, a_{\tilde d+1}$ denote the angular coordinates of the
$(\tilde d+1)$-sphere. The functions $E$ and $\tilde E$ are also
assumed to depend only on $r$. The NS-NS 3-form responsible for the
off-diagonal component of the metric is separately written as
\bea
H_{uij} = \pa_{[i} b_{j]},
\label{nsns}
\ena
such that it satisfies the Bianchi identity. Here the indices $i,j$ denote
the directions transverse to the branes ($z_i, z_j$ or $r$ and angles).

Solving for the dilaton in eq.~(\ref{dila}) gives
\bea
\phi'' + \frac{(\tilde d+1)}{r}\phi'  = -\shalf \sum_A\e_A a_A S_A(E'_A)^2,
\label{dile}
\ena
where $S_A$ is defined as
\bea
S_A = \exp\Big({\e_A a_A\phi - 2\sum_{\a \in q_A} u_\a}\Big).
\label{sa}
\ena
The sum of $\a$ runs over the world-volume of the branes
( $u,v$ and $(q_A-1)\; y^\a$ coordinates, and so $\sum_{\a \in q_A} u_\a
=2u_0 + \sum_{\a=2}^{q_A} u_\a$ for example).

The Ricci tensors for the metric~\p{met} are summarized in the appendix.
The field eqs.~(\ref{Einstein}) are simplified considerably by
imposing the condition
\bea
2u_0+ \sum_{\a=2}^{d-1} u_\a +\tilde d B = 0.
\label{conde}
\ena
It is known that under this condition, all the known intersecting
brane solutions can be derived~\cite{NO}-\cite{aeh}. We can see that
this condition makes most of the field equations in \p{Einstein} linear,
allowing superpositions of obtained solutions, and it is closely related
to BPS conditions. Thus it is also expected to be sufficient to impose this
condition in deriving BPS brane solutions in the pp-wave background.
We will relax this restriction in our search for non-extremal solutions
in the next section.

Using the condition~\p{conde}, we can write eqs.~\p{Einstein} in
the following form:
\bea
&& u_0'' +  \frac{(\tilde d+1)}{r}u_0'  = \sum_A
\frac{D-q_A-3}{2(D-2)} S_A(E'_A)^2,
\label{uve}
\\
&&\Bigg[u_0'' + \frac{(\tilde d+1)}
{r}u_0' + \shalf K^{-1}\Big(\Box^{(\tilde d + 2)}
+ \sum_\a e^{2(B-u_\a)}\pa^2_\a\Big) K\Bigg] \nn
&& \hs{20} = \sum_A\frac{D-q_A-3}{2(D-2)} S_A(E'_A)^2
- \frac{1}{4} K^{-1} e^{-(2u_0+2B +\phi)}(\pa_{[i} b_{j]})^2,
\label{uue}
\\
&& u_\a'' + \frac{(\tilde d+1)}{r}u_\a' = \sum_A
\frac{\d^{(\a)}_A}{2(D-2)} S_A(E'_A)^2,
\label{albee}
\\
&&B'' + \tilde d (B')^2 + \frac{(\tilde d+1)}{r} B'
+ 2(u_0')^2 +\sum_{\a=2}^{d-1} (u_\a')^2 \nn
&&\hs{50}
= - \shalf (\phi')^2 + \sum_A\frac{D-q_A-3}{2(D-2)}S_A(E'_A)^2,
\label{rre}
\\
&& B'' -\frac{1}{r^2}+\frac{(\tilde d+1)}{r}\Big(B' + \frac{1}{r}\Big)
- \frac{\tilde d }{r^2}
= - \sum_A\frac{q_A+1}{2(D-2)}S_A(E'_A)^2,
\label{abe}
\ena
where $\d_A^{(\a)}$ is defined as
\bea
\d_A^{(\a)} = \left\{ \begin{array}{l}
D-q_A-3 \\
-(q_A+1)
\end{array}
\right.
\hs{5}
{\rm for} \hs{3}
\left\{
\begin{array}{l}
y_\a \mbox{   belonging  to $q_A$-brane} \\
{\rm otherwise}
\end{array}
\right. ,
\ena
and $\e_A= +1 (-1)$ is for electric (magnetic) backgrounds.
The equations (\ref{uve}), (\ref{uue}), (\ref{albee}), (\ref{rre}) and
(\ref{abe}) are the $uv, uu, \a\b, rr$ and $ab$ components of the
Einstein equations (\ref{Einstein}) respectively.

The field equation for the NS-NS 3-form~\p{nsns} then leads to
\bea
\pa_{[i}b_{j]} = e^{2u_0 + 2B+ \phi} \mu_{ij},
\label{nsns3}
\ena
where $\mu_{ij}$ is constant.

{}From eqs.~(\ref{uve}), (\ref{uue}) and \p{nsns3}, we get the following
differential equation for $K$:
\bea
\Box^{(\tilde d + 2)} K +  \sum_\a e^{2(B-u_\a)}\pa^2_\a K
= \frac12 e^{2u_0 +2B +\phi}(\mu_{ij})^2,
\label{ka}
\ena
where $\Box^{(\tilde d + 2)}$ is the Laplacian in flat $\tilde d + 2$
dimensional spacetime.

Solving eq.~(\ref{field}) for the field strength gives
\bea
S_A E'_A = c_A r^{-(\tilde d+1)}.
\label{fielde}
\ena
Using this, one can integrate eqs.~(\ref{dile}), (\ref{uve}),
(\ref{albee}) and (\ref{abe}) to obtain a set of first order equations
for $\phi$, $u_0$, $u_\a$ and $B$. They are
\bea
\label{dilae}
\phi' &=& -\sum_A \frac{\e_A a_A c_A}{2} \frac{E_A}{r^{\tilde d+1}}
+ \frac{c_\phi}{r^{\tilde d+1}}, \\
\label{uv1e}
u'_0 &=& \sum_A \frac{D-q_A-3}{2(D-2)} c_A \frac{E_A}{r^{\tilde d+1}}
+ \frac{c_0}{r^{\tilde d+1}}, \\
\label{albe1e}
u'_\a &=& \sum_A  \frac{\d^{(\a)}_A}{2(D-2)} c_A
\frac{E_A}{r^{\tilde d+1}} + \frac{c_\a}{r^{\tilde d+1}}, \\
B' &=& - \sum_A \frac{q_A+1}{2(D-2)} c_A \frac{E_A}{r^{\tilde d+1}}
+ \frac{c_b}{r^{\tilde d+1}}.
\label{ab1e}
\ena

Putting $u'_0, u'_\a, B'$ into eq.~(\ref{rre}) and equating
$E_A$-independent part equal to zero, we get
\bea
c_\phi = c_0 = c_\a = c_b = 0,
\ena
and $E_A$-dependent part equal to zero, we get
\bea
\sum_{A,B}\Bigg[M_{AB}\frac{c_A}{2} + r^{\tilde d+1}\Big(\frac{1}{E_A}\Big)'
\d_{AB}\Bigg]\frac{c_B}{2}E_A E_B = 0,
\label{int}
\ena
where
\bea
M_{AB} = \tilde d \frac{(q_A + 1)(q_B +1)}{{(D-2)}^2} + \sum_{\a=0}^{d-1}
\frac{\d^{(\a)}_A\d^{(\a)}_B}{{(D-2)}^2} + {1\over 2} \e_A a_A \e_B a_B.
\ena
$M_{AB}$ being constant, eq. (\ref{int}) cannot be satisfied for
arbitrary function $E_A$ unless the second term inside the square
bracket is a constant. This gives
\bea
E_A = N_A H^{-1}_A,
\label{harm}
\ena
where the function $H_A$ is defined as
\bea
H_A = 1 + \frac{Q_A}{r^{\tilde d}}.
\label{hae}
\ena

Putting $A=B$ in eq.~(\ref{int}), we get
\bea
\frac{c_A}{2} = \frac{\tilde d Q_A}{N_A M_{AA}}\equiv \frac{\tilde d
  Q_A}{N_A}\frac{D-2}{\Delta_A},
\label{ca}
\ena
where $\Delta_A$ is given by
\bea
\Delta_A = (q_A + 1) (D-q_A-3) + \shalf a_A^2 (D-2).
\label{Delta}
\ena

Integrating eqs.~(\ref{dilae}), (\ref{uv1e}), (\ref{albe1e}) and
(\ref{ab1e}), we get
\bea
\phi &=& \sum_{A} \e_A a_A \frac{D-2}{\Delta_A}  \ln H_A, \nn
u_0 &=& - \sum_{A} \frac{D-q_A-3}{\Delta_A} \ln H_A, \nn
u_\a &=& - \sum_{A} \frac{\d_{A}^{(\a)}}{\Delta_A} \ln H_A, \nn
B &=& \sum_{A} \frac{q_A+1}{\Delta_A} \ln H_A,
\label{warpe}
\ena
where the integration constants are put equal to zero by the
requirement that asymptotically the warp factors approach to $1$.

Using eqs.~(\ref{warpe}), one can write down the expression
(\ref{sa}) for $S_A$ as
\bea
S_A= H^2_A.
\ena
By use of eqs.~(\ref{warpe}), eq.~(\ref{ka}) for $K$ becomes
\bea
\Big(\Box^{(\tilde d + 2)}  + \sum_{\a=2}^{d-1} \prod_A
H_A^{2\frac{\c_A^{(\a)}}{\Delta_A}} \pa^2_\a \Big) K
= - \frac12 (\mu_{ij})^2 \prod_A H_A^{l_A},
\label{ka1}
\ena
where we have defined
\bea
\c_A^{(\a)} = \left\{ \begin{array}{l}
D-2 \\
0
\end{array}
\right.
\hs{5}
{\rm for} \hs{3}
\left\{
\begin{array}{l}
y_\a \mbox{   belonging  to $q_A$-brane} \\
{\rm otherwise}
\end{array},
\right.
\label{gamma}
\ena
and
\bea
l_A = \frac{4(q_A+1)+ \e_A a_A(D-2)-2(D-2)}{\Delta_A}.
\ena
We note that for D-branes in $D=10$, $l=0$ and $\frac{2(D-2)}{\Delta_A}=1$.
For a single D$q_A$-brane, eq.~(\ref{ka1}) admits a solution of the form
\bea
K = c + \frac{\mathcal Q}{r^{\tilde d}} -\frac{(\mu_{ij})^2}{32}
\Big(r^2  + \sum_{\a}y_\a^2 + \frac{(q_A-1)}{(\tilde d-2)}\frac{Q_A}
{r^{\tilde d -2}}\Big),
  ~~~({\rm for}~~\tilde d \ne 2)
\label{ka2}
\ena
and
\bea
K = c + \frac{\mathcal Q}{r^{\tilde d}} -\frac{(\mu_{ij})^2}{32}
\Big(r^2 + \sum_{\a}y_\a^2 - (q_A-1)Q_A\ln r\Big),
 ~~~({\rm for }~~ \tilde d = 2)
\label{ka3}
\ena

Now, using eqs.~(\ref{fielde}) and (\ref{ca}), we can determine
the normalization constant $N_A$ as
\bea
N_A = \sqrt{\frac{2(D-2)}{ \Delta_A}}.
\ena

Our metric and background fields are thus finally given by, after
putting all the warp factors etc. we obtained by solving the
Einstein equations,
\bea
ds_D^2 &=& \prod_A H_A^{2 \frac{q_A+1}{\Delta_A}} \Bigg[ \prod_A
 H_A^{- 2 \frac{D-2}{\Delta_A}} \Big\{ - 2dudv + K du^2\Big\} \nn
&& \hs{15} + \; \sum_{\a=2}^{d-1}\prod_A H_A^{- 2 \frac{\c_A^{(\a)}}{\Delta_A}}
dy_\a^2 + dr^2 + r^2 d\Omega_{\tilde d+1}^2 \Bigg], \nn
&& E_A = \sqrt{\frac{2(D-2)}{ \Delta_A}} H^{-1}_A.
\ena
where $\c_A^{(\a)}$ is defined in \p{gamma} and the function $K$ is
in eq.~(\ref{ka1}).

For $A\ne B$, we have $M_{AB}=0$ from eq.~(\ref{int}). This gives the
intersection rules for the branes. If $q_A$-brane and $q_B$-brane
intersect over ${\bar q} (\leq q_A, q_B)$ dimensions, this gives
\bea
{\bar q} = \frac{(q_A+1)(q_B+1)}{D-2}-1 - \shalf \e_A a_A \e_B a_B.
\label{ints}
\ena
For D-branes
\bea
\e_A a_A = \frac{3-q_A}{2},
\ena
and we get
\bea
\bar q=\frac{q_A+q_B}{2} -2.
\label{drule}
\ena
The results presented here are the pp-wave generalization of the
intersection rules already discussed in the literature
\cite{NO}-\cite{aeh}. The amount of supersymmetry preserved by
these brane configurations can be obtained by solving the Killing spinor
equations explicitly. In the present case, the lightcone directions
are lying along the brane, whereas the other pp-wave directions
are transverse to the brane world-volume. The supersymmetric properties
of such D-brane configurations and their bound states are already
discussed in~\cite{kamal,kamal1} for $D=10$. For the D-branes in
the background under consideration, there always exist 16 `standard'
Killing spinors $(\epsilon_{\pm})$ satisfying
$\Gamma^{\hat u}\epsilon_{\pm} = 0$ \cite{pope,stelle}. The rest of the
supersymmetry preservation depends on the number of solutions to the condition
$(\pa_{\hat i} b_{\hat j})\Gamma^{\hat i \hat j}\e_\pm=0$ and the standard
D-brane supersymmetry conditions. For the special case of $H_{u12}=H_{u34}$,
the condition $(\pa_{\hat i} b_{\hat j})\Gamma^{\hat i \hat j}\e_\pm=0$ breaks
half of the supersymmetry and the D-brane configurations in this background
preserve $1/8$ supersymmetries \cite{kamal,kamal1}.

\section{Non-Extremal Solutions}

In this section, we present the non-extremal generalization of
the solutions analyzed in the previous section. This could be done
by directly starting with a metric ansatz with blackening functions along
with the arbitrary warp factors and then solving the field equations to
fix each of them accordingly. We follow a slightly different approach.
Instead of putting the blackening functions directly into the metric ansatz,
without any loss of generality, we shall deform the condition (\ref{conde}) as
\bea
2u_0+ \sum_{\a=2}^{d-1} u_\a +\tilde d B = \ln g,
\label{condn}
\ena
where $g$ being a function of $r$ only.

In our derivation of extremal solutions in the preceding section,
we have set the above combination to zero. This is sufficient to
find extremal solutions. Here we relax this restriction and search
for general solutions.

Using (\ref{condn}), eqs.~(\ref{Einstein}) can be rewritten as
\bea
&&\Bigg[ u_0'' + \Big(\frac{g'}{g} + \frac{(\tilde d+1)}{r}\Big)u_0' \Bigg]
= \sum_A\frac{D-q_A-3}{2(D-2)} S_A(E'_A)^2,
\label{uvn}
\\
&&\Bigg[u_0'' + \Big(\frac{g'}{g} + \frac{(\tilde d+1)}{r}\Big)u_0' 
+\shalf \frac{g'}{g} \frac{K'}{K}
+ \shalf K^{-1}\Big(\Box^{(\tilde d + 2)}
+ \sum_\a e^{2(B-u_\a)}\pa^2_\a\Big) K\Bigg] \nn
&& \hs{25} = \sum_A\frac{D-q_A-3}{2(D-2)} S_A(E'_A)^2
- \frac{1}{4} K^{-1} e^{-(2u_0+2B +\phi)} (\pa_{[i} b_{j]})^2
\label{uun}
\\
&&\Bigg[ u_\a'' + \Big(\frac{g'}{g} +\frac{(\tilde d+1)}{r}\Big)u_\a'
\Bigg] \d_{\a\b} = \sum_A \frac{\d^{(\a)}_A}{2(D-2)} S_A(E'_A)^2,
\label{alben}
\\
&&\Bigg[B'' + \tilde d (B')^2 + \Big(-\frac{g'}{g} + \frac{(\tilde d+1)}{r}
\Big)B' + \Big(\frac{g'}{g}\Big)^{'} + 2(u_0')^2
+\sum_{\a=2}^{d-1} (u_\a')^2\Bigg]\nn
&&\hs{50}= - \shalf (\phi')^2 + \sum_A\frac{D-q_A-3}{2(D-2)}S_A(E'_A)^2,
\label{rrn}
\\
&&\Bigg[ B'' -\frac{1}{r^2}+ \Big( \frac{g'}{g} + \frac{(\tilde d+1)}{r}\Big)
\Big(B' + \frac{1}{r}\Big) - \frac{\tilde d}{r^2}\Bigg]
= - \sum_A\frac{q_A+1}{2(D-2)}S_A(E'_A)^2,
\label{abn}
\ena
where $S_A$ is defined as
\vs{-1}
\bea
S_A = \exp\Big({\e_A a_A\phi - 2\sum_{\a \in q_A} u_\a}\Big)
\label{san}
\ena
Similarly the dilaton eq.~(\ref{dila}) can be written as
\bea
\phi'' +  \Big(\frac{g'}{g}+\frac{(\tilde d+1)}{r}\Big) \phi'
= -\shalf \sum_A\e_a a_A S_A(E'_A)^2,
\label{diln}
\ena

Solving the equation for the field strengths gives
\bea
S_A E'_A = c_A g^{-1}r^{-(\tilde d + 1)}.
\label{fieldn}
\ena

To solve for $g(r)$, we multiply eqs.~(\ref{uvn}) and ~(\ref{abn}) by $2$ and
$\tilde d$ respectively and add to eq.~(\ref{alben}). Using the
condition (\ref{condn}), we get
\bea
g(r) = 1 - \Big(\frac{r_0}{r}\Big)^{2\tilde d}.
\ena

Solving the dilaton eq.~({\ref{diln}}) and eqs.~(\ref{uvn}), (\ref{alben})
and (\ref{abn}) by using the solution
(\ref{fieldn}), we get the following set of first order eqs.
\bea
\label{dilan}
\phi' &=& -\sum_A \frac{\e_A a_A c_A}{2} E_A h^{-1} + c_\phi h^{-1}, \\
\label{uv1n}
u'_0 &=& \sum_A \frac{D-q_A-3}{2(D-2)} c_A E_A h^{-1} + c_0 h^{-1},\\
\label{albe1n}
u'_\a &=& \sum_A  \frac{\d^{(\a)}_A}{2(D-2)} c_A E_A h^{-1} + c_\a h^{-1},\\
B' &=& - \sum_A \frac{q_A+1}{2(D-2)} c_A E_A h^{-1} +
\frac{1}{\tilde d} \frac{g'}{g} + c_b h^{-1},
\label{ab1n}
\ena
where $h$ is defined as
\bea
h = g r^{\tilde d + 1}.
\ena
Putting values of $u'_0, u'_\a, B'$ into eq.~(\ref{rrn}) and equating
the $E_A$-independent part to zero, we get the condition
\bea
2 c^2_0 +\sum_{\a=2}^{d-1} c^2_\a + \tilde d c^2_b + \shalf c^2_\phi
= 4 \tilde d(\tilde d +1)r^{2\tilde d}_0,
\label{cond2n}
\ena
and similarly, $E_A$-dependent part equal to zero gives
\bea
\sum_{A,B}\Bigg[M_{AB}\frac{c_A}{2} + \Big(h\Big(\frac{1}{E_A}\Big)'
+\frac{\tilde c_A}{E_A}\Big)\d_{AB}\Bigg]\frac{c_B}{2}E_A E_B = 0,
\label{intn}
\ena
where
\bea
M_{AB} = \tilde d \frac{(q_A + 1)(q_B +1)}{{(D-2)}^2} + \sum_{\a=0}
\frac{\d^{(\a)}_A \d^{(\a)}_B}{{(D-2)}^2} + {1\over 2} \e_A a_A \e_B a_B,
\ena
and
\bea
\tilde c_A = -2 \tilde d c_b\frac{q_A + 1}{D-2} + 2\sum_{\a=0}^{d-1}
\frac{\d^{(\a)}_A}{D-2}c_\a - \e_A a_A c_\phi.
\label{tca}
\ena
$M_{AB}$ being constant, eq.~(\ref{int}) cannot be satisfied for
arbitrary function $E_A$ unless the second term inside the square
bracket is a constant. This gives
\bea
E_A = \frac{N_A }{1 - \b_A(1 - f^{-\a_A})},
\ena
where
\bea
\a_A= \frac{\tilde c_A}{2\tilde d r^{\tilde d}_0},
\label{alpha}
\ena
$\b_A$ and $N_A$ are constants and the function $f(r)$ is defined as
\bea
f(r) = \frac{1 - \Big(\frac{r_0}{r}\Big)^{\tilde d}}{1 +
  \Big(\frac{r_0}{r}\Big)^{\tilde d}}.
\ena
Putting $A=B$ in eq.~(\ref{intn}), we get
\bea
\frac{c_A}{2} = \frac{\tilde c_A(\b_A-1)}{N_A M_{AA}}
\equiv \frac{\tilde c_A(\b_A-1)}{N_A}\frac{D-2}{\Delta_A},
\label{can}
\ena
where $\Delta_A$ is given in \p{Delta}.

Integrating eqs.~(\ref{dilan}), (\ref{uv1n}), (\ref{albe1n}) and
(\ref{ab1n}), we get
\bea
\phi &=& \sum_{A} \e_A a_A  \frac{D-2}{\Delta_A}  \ln \tilde H_A
+ \frac{c_\phi}{2\tilde d r^{\tilde d}_0}\ln f, \nn
u_0 &=&  -\sum_{A} \frac{D-q_A-3} {\Delta_A} \ln \tilde H_A
+\frac{c_0}{2\tilde d r^{\tilde d}_0}\ln f, \nn
u_\a &=& -\sum_{A} \frac{\d^{(\a)}_A} {\Delta_A} \ln \tilde H_A
+\frac{c_\a}{2\tilde d r^{\tilde d}_0}\ln f, \nn
B &=&  \sum_{A} \frac{q_A+1}{\Delta_A} \ln \tilde H_A
+ \frac{1}{\tilde d}\ln g +\frac{c_b}{2\tilde d r^{\tilde d}_0}\ln f,
\label{warpn}
\ena
where $\tilde H_A$ is given by
\bea
\tilde H_A =N_A E_A^{-1} f^{\a_A}
= \left\{1-\b_A (1-f^{-\a_A})\right\}f^{\a_A},
\label{han}
\ena
and  the integration constants are fixed by the
requirement that asymptotically the warp factors approach to $1$.

Using eqs.~(\ref{warpn}), one can write down the expression
(\ref{san}) for $S_A$ as
\bea
S_A= N_A^2 E^{-2}_A f^{\a_A}.
\ena
Now, using eqs.~(\ref{fieldn}) and (\ref{can}), we can determine
the normalization constants $N_A$ as
\bea
N_A = \sqrt{\frac{2(\b_A-1)}{\b_A}\frac{(D-2)}{\Delta_A}}.
\ena
We also have
\bea
2c_0+\sum_{\a=2}^{d-1} c_\a + \tilde d c_b =0,
\label{condnn}
\ena
{}from the relation~\p{condn}. By use of this relation, $\tilde c_A$ in
eq.~\p{tca} can also be written as
\bea
\tilde c_A = 2 \sum_{\a \in q_A} c_\a - \e_A a_A c_\phi.
\ena

Our metric and background fields are thus finally given by, after
putting all the warp factors etc. that we get by solving the
Einstein equations,
\bea
ds_D^2 &=& \prod_A \tilde {H}_A^{2 \frac{q_A+1}{\Delta_A}} \Bigg[  \prod_A
\tilde{H}_A^{ - 2 \frac{D-2}{\Delta_A}} f^{{c_0}/{\tilde d r_0^{\tilde d}} }
\Big\{-2dudv+ K du^2\Big\} \nn
&& \hs{-10}+\; \sum_{\a=2}^{d-1}\prod_A \tilde H_A^{- 2 \frac{\c_A^{(\a)}}
{\Delta_A}} f^{{c_\a}/{\tilde d r_0^{\tilde d}}} dy_\a^2
+ f^{{c_b}/{\tilde d r_0^{\tilde d}}}
g^{2/{\tilde d}} \Big(dr^2 + r^2 d\Omega_{\tilde d+1}^2\Big) \Bigg], \nn
&& E_A =  \sqrt{\frac{2(\b_A-1)}{\b_A}\frac{(D-2)}{ \Delta_A}}
\tilde {H}^{-1}_A f^{\a_A}.
\label{metn}
\ena
where the function $K$ is given from eq.~(\ref{uvn}) and ~(\ref{uun}) by
\bea
\hs{-5}
\Big[ \Box^{(\tilde d+2)} + g^{-1} \partial_r g \partial_r + \sum_{\a=2}^{d-1} g^{2/\tilde d}
f^{(c_b-c_\a)/\tilde d r_0^{\tilde d}} \prod_A \tilde H_A^{2
\frac{\c_A^{(\a)}}{\Delta_A}} \pa_\a^2 \Big] K \cr
&\cr
= - \frac{(\mu_{ij})}{2}^2 g^{2/\tilde d} f^{(c_0+c_b+c_\phi/2)/\tilde d
r_0^{\tilde d}} \prod_A \tilde H_A^{l_A}.
\label{ka4}
\ena

The extremal limit corresponds to $r_0 \to 0$ and $\b_A \to \infty$,
keeping $\b_A r_0^{\tilde d}$ finite. In this limit, noting that
$\a_A$ is finite from \p{cond2n}, \p{tca} and \p{alpha}, we find
\bea
\tilde H_A \to 1+2 \a_A \b_A \Big(\frac{r_0}{r}\Big)^{\tilde d},
\ena
so that $2\a_A\b_Ar^{\tilde d}_0$ is a parameter corresponding to the
charge $Q_A$ in the extremal solution.

For $A\ne B$, we have $M_{AB}=0$ from eq.~(\ref{intn}). This gives the
intersection rules for the branes. If $q_A$-brane and $q_B$-brane
intersect over ${\bar q} (\leq q_A, q_B)$ dimensions, this gives
\bea
{\bar q} = \frac{(q_A+1)(q_B+1)}{D-2}-1 - \shalf \e_A a_A \e_B a_B.
\label{intsn}
\ena
We also get the rule \p{drule} for D-branes.
We, once again, would like to point out that the above analysis is
consistent with the flat space analysis performed in
\cite{NO}-\cite{aeh}. These branes are nonsupersymmetric for
arbitrary values of non-extremal parameter $r_0$. One can check that
in the limit $r_0\rightarrow 0$, both $f$ and $g$ goes to one and we get
back to the supersymmetric solutions presented in the previous section.

\section{Black branes and Horizons}

Non-extremal D-brane solutions usually admit horizons and are known as
black branes. Though this is certainly true in flat spacetime, not all
non-extremal solutions in pp-wave admit regular horizon~\cite{liu,bre}.
In particular, if the background admits a null Killing isometry, the
corresponding brane solutions are found to have naked singularities.
While the authors of \cite{liu} have analyzed pp-wave background
supported by 5-form RR fields, our backgrounds have different isometries
and it would be interesting to examine if one can find some black branes in
this background. Also since the construction of the general solutions in the
preceding sections is somewhat abstract, it would be instructive to present
explicit examples of D-branes in the background (\ref{plane}) and discuss
properties of their horizons.

First let us review the criteria developed in ref.~\cite{liu} for the
existence of horizon in pp-wave spacetimes. For the black holes in
spacetimes which are not asymptotically flat e.g. pp-waves in our context,
no rigorous definition of event horizon exists, partly because of difficulty
in identifying the future null infinity. So one has to content with
the working hypothesis that a black hole is the region of the spacetime
causally disconnected from the asymptotic infinity. Thus the coordinate time
along timelike or null geodesics to reach the asymptotic infinity is
arbitrarily large. PP-wave spacetime admits two Killing vectors
$\frac{\pa}{\pa u}~ {\rm and}~ \frac{\pa}{\pa v}$.
For $ \frac{\pa}{\pa u}$ timelike, coordinate time can be
identified with $u$ and the following criteria for the existence of
the horizon will apply.

Suppose the warp factors behave in the vicinity of the horizon
at $r_0$ as
\bea
e^{2u_0}\sim (r-r_0)^{2a}, ~~~~~~e^{2B}\sim (r-r_0)^{2b}, ~~~~~~
K\sim (r-r_0)^{2h} ~{\rm sgn}~ K.
\ena
Then the conditions for the existence of the horizon are:
\bea
&&a+b>-1, ~~~~~~~~~~~~ a-b\ge 1 ~~~~~~~~~~~~{\rm if}~h\ge 0  \nn
&&a+b +|h|>-1, ~~~~a-b\ge 1 + |h| ~~~~~~{\rm if} ~h< 0 ~~{\rm and}~~
{\rm sgn}~K = + \nn
&&a+b>-1, ~~~~~~~~~~~~ a-b\ge 1 ~~~~~~~~~~~~{\rm if} ~K\sim |\ln(r-r_0)|.
\label{condh}
\ena
For $\frac{\pa}{\pa u}$ spacelike or null, one has to choose the
coordinate time as some linear combination of $u$ and $v$ and second
condition is replaced by
\bea
a+b +|h|>-1, ~~~~a-b\ge 1 - |h| ~~~~~~{\rm if} ~h< 0 ~~{\rm and}~~
{\rm sgn}~K = +.
\label{cri1}
\ena

For the pp-wave spacetime which are asymptotic to BFHP plane wave~\cite{bfhp}
(as is the case with our solutions), $\frac{\pa}{\pa u}$ is always
timelike and one can use the first set of conditions given above.
In this case, it should be noted that condition with $h<0$ is stronger
than the condition with $h\ge 0$ in the sense that if there exists no
horizon for $h\ge0$, and then it cannot exist for $h<0$ either. Since our
solutions are asymptotic to plane wave spacetime, we do not have to
worry about the second set of conditions.

However this criterion (hereafter referred to as the first criterion),
even when it is satisfied, might not be sufficient to guarantee the
existence of a horizon~\cite{bre}. It is not difficult to
see that all scalar curvature invariants in the background ~(\ref{met})
behave regularly in the near horizon limit. However, there may be divergence
in the Riemann tensors $R_{\mu\nu\rho}{}^\sigma$~\cite{bre}. Now the
geodesic deviation equation for a test particle moving in the spacetime
(\ref{met}) is given by
\bea
\frac{D^2 x^\mu}{d\tau^2} = - R^\mu{}_{\nu\rho\sigma} u^\nu x^\rho u^\sigma
\ena
where $x^\mu (\tau)$ and $u^\mu$ are the displacement vector and four
velocity of the test particle respectively. Hence the geodesic deviation
equation may become singular and an observer traveling along a causal
timelike or null geodesics will feel infinite tidal forces. The relative
motion has an invariant meaning only in an orthonormal frame (with basis
vectors ${e_a}$ obeying
$e_a \cdot e_b = \eta_{ab}$). The geodesic equation in this frame becomes
\bea
\ddot{x^i} = - R^i{}_{0j0}x^j,
\ena
where $x^i= e^i_\mu x^\mu$ etc. The most natural choice for the
orthonormal frame is the parallel transported frame and we shall see,
by some examples, that the Riemann tensors as measured in the parallel
transported frame diverges for all these cases, in contrast to the first
criterion mentioned above. This suggests that the first criterion is only
a necessary condition but not sufficient.

For simplicity, we set the parameters $\a, \b, \c$ and $\d$ as
\bea
\a = \frac{c_0}{2 \tilde d r_0^{\tilde d}}, \quad
\b = \frac{c_2}{2 \tilde d r_0^{\tilde d}}
= \frac{c_3}{2 \tilde d r_0^{\tilde d}} = \cdots, \quad
\c = \frac{c_b}{2 \tilde d r_b^{\tilde d}}, \quad
\d = \frac{c_\phi}{2 \tilde d r_0^{\tilde d}}.
\ena
Also since we are interested in the near-horizon limit of warp factors,
it is useful to note the behaviour of functions $\tilde H_A, ~f$ and $g$
in this limit:
\bea
\tilde H_A \sim (r-r_0)^{-\frac{|\a_A|}{2}+\frac{\a_A}{2}}, ~~~~~
f\sim (r-r_0), ~~~~~g\sim(r-r_0).
\ena

\subsection{D3-brane}

First we consider the metric for the extremal D3-brane solution
\bea
ds^2 &=& H^{-\shalf}_3\left[-2dudv + K du^2 + dy_2^2+ dy_3^2\right]
 + H^{\shalf}_3\left[dz_1^2 + \cdots + dz_6^2 \right],\nn
&& H_{uij} = \mu_{ij}.
\ena
The corresponding eq.~(\ref{ka1}) for $K$ becomes
\bea
\Big( \Box^{(6)}  + H_3 (\pa^2_{y_2}+ \pa^2_{y_3})\Big)K
= - \frac{(\mu_{ij})^2}{2},
\label{kae}
\ena
where $\Box^{(6)}$ is the Laplacian in the six-dimensional transverse space
spanned by $z_1, ..., z_6$ and $H_A$ is given by eq.~(\ref{hae})
with $\tilde d = 4$.

Equation~(\ref{kae}) admits a solution of the form
\bea
K = c + \frac{\mathcal Q}{r^4} -\frac{1}{32}(\mu_{ij})^2 \Big(r^2 +
y_2^2+y_3^2 + \frac{Q_3}{r^2}\Big).
\ena
So for the extremal D3-brane solution $r_0=0$, $a=-b=1$ and $h=-2$ and
the criteria for the existence of the horizon is not satisfied.

Next we consider the non-extremal deformation of the above
solution. The metric of the non-extremal solution is given by
\bea
ds^2 &=& \tilde {H}^{-\shalf}_3f^{2\a}(-2dudv + K du^2)
+ \tilde H_3^{-\shalf} f^{2\b}(dy_2^2+ dy_3^2) \nn
&& \hs{10} +\; \tilde{H}^{\shalf}_3f^{2\c}
g^\shalf (dz_1^2 + \cdots + dz_6^2 ),
\label{d3}
\ena
where $\tilde H_3$ is given by eq.~(\ref{han}) with $\tilde d = 4$.
The corresponding eq.~(\ref{ka4}) for $K$ then becomes
\bea
\Big(\Box^{(6)} + g^{-1} \partial_r g \partial_r + g^\shalf f^{2(\c-\b)}\tilde{H}_3 (\pa^2_{y_2}+ \pa^2_{y_3})
\Big)K = - \frac{(\mu_{ij})^2}{2} g^{\shalf}f^{2\a+2\c +\d}.
\label{kan}
\ena
The parameters $\a, \b $ and $\c$ satisfy the relation
\bea
\a+\b+2\c = 0,
\label{cond4n}
\ena
and the relation (\ref{cond2n}) becomes
\bea
8\a^2 + 24\c^2 +16\a\c + \d^2 = \frac{5}{2}.
\label{cond5n}
\ena
Using these relations, we find from \p{d3}
\bea
a=\frac{|\a_{\rm D3}|-\a_{\rm D3}}{8}+\a, \quad
b=-\frac{|\a_{\rm D3}|-\a_{\rm D3}}{8}+\c+\frac{1}{4}, \quad
\a_{\rm D3} = -8\c.
\label{cri2}
\ena

It is not difficult to see that the relations \p{condh}, \p{cond4n}-\p{cri2}
cannot be satisfied for real values of parameters $\a, \c,$ and $\d$.
So the non-extremal deformation of the D3-brane solution does not admit
horizon.

Now we would like to check how the curvature tensors behave in the near
horizon limit. As already mentioned, the possible divergent quantity is
the Riemann tensor measured in an orthonormal frame. A natural choice for
it is the parallel transported frame as emphasized in \cite{bre}.
The Riemann tensors in the parallel transported frame are given by
\bea
&& R_{tptp}\equiv R_{\mu\nu\rho\sigma} t^\mu p^\nu t^\rho p ^\sigma, \nn
&& R_{tpti}\equiv R_{\mu\nu\rho\sigma} t^\mu p^\nu t^\rho n_i^\sigma, \nn
&& R_{titj}\equiv R_{\mu\nu\rho\sigma} t^\mu n_i^\nu t^\rho n_j^\sigma,
\ena
where $t^\mu,p^\mu,n_i^\mu,n_j^\mu$ are the unit vectors in the parallel
transported frame.

Explicit form of these vectors for D3-brane can be read off from
ref.~\cite{bre}. Using the expression for Riemann tensors given in
the appendix, one can see that $R_{tptp}\sim (K,_{rr} + 3 \frac{K,_r}{r}
\cos^2({\tau})), R_{tpti}\sim \frac{1}{r^2} K,_{rx}\cos({\tau})$ and
$R_{titj}\sim \frac{1}{r^4}(K,_{y_iy_j} + \d_{ij} r^3 K,_r)$ for the
extremal solution, and they diverge as $\frac{1}{r^6}, 1$ and
$\frac{1}{r^4}$ for $i \ne j$ and $\frac{1}{r^6}$ for $i = j$, respectively.
Similarly we can see that some components of the Riemann tensors behave as
$\frac{1}{(r-r_0)^a}$ with $a >0$ for non-extremal solutions and hence
are divergent in near horizon limit. This means that there is no regular
horizon in these extremal and non-extremal solutions, in agreement with
the above conclusion.

\subsection{D3-D3 system}

Next let us consider the extremal D3-D3 system. The metric is given by
\bea
ds^2 &=& H_3^{-\frac{1}{2}} H_{3'}^{-\frac{1}{2}}(-2dudv + K du^2)
 + H_3^{-\frac{1}{2}}H_{3'}^{\frac{1}{2}} (dy^2_2 +dy_3^2) \nn
&& \hs{5}+\; H_3^{\frac{1}{2}}H_{3'}^{-\frac{1}{2}}(dy^2_4+dy^2_5)
+ H_3^{\frac{1}{2}} H_{3'}^{\frac{1}{2}} (dz_1^2 + \cdots + dz_4^2),\nn
&& H_{uij} = \mu_{ij}.
\ena
The corresponding eq.~(\ref{ka4}) for $K$ then becomes
\bea
[\Box^{(4)}  + H_3 (\pa_{y_2}^2 +\pa^2_{y_3})
+ H_{3'} (\pa^2_{y_4} + \pa^2_{y_5} )]K= - \shalf (\mu_{ij})^2,
\label{kaed33}
\ena
where $\Box^{(4)}$ is again the Laplacian in the four-dimensional
transverse space spanned by $z_1, \cdots, z_4$ and $H_3$ and $H_{3'}$ are
given by eq.~(\ref{hae}) with $\tilde d = 2$.

Equation~(\ref{kaed33}) admits a solution of the form
\bea
K = c + \frac{\mathcal Q}{r^2} -\frac{1}{32}(\mu_{ij})^2\Big(r^2 +
y_2^2+y_3^2 +y_4^2+y_5^2 - 2 Q_3\ln r - 2 Q_{3'}\ln r\Big).
\ena
So for the extremal D3-D3 system $a=-b=1$ and $h=-1$ and  the criteria
for the existence of the horizon is satisfied. However, we shall see that
this is not sufficient for the existence of the regular horizon.

Now let us consider the non-extremal deformation of the D3-D3 system.
The metric is given by
\bea
ds^2 &=& H_3^{-\frac{1}{2}} H_{3'}^{-\frac{1}{2}}f^{2\a}(-2dudv + K du^2)
 + H_3^{-\frac{1}{2}}H_{3'}^{\frac{1}{2}} f^{2\b}(dy^2_2 +dy_3^2) \nn
&& \hs{5}+\; H_3^{\frac{1}{2}}H_{3'}^{-\frac{1}{2}}f^{2\b}(dy^2_4+dy^2_5)
+ H_3^{\frac{1}{2}} H_{3'}^{\frac{1}{2}} f^{2\c}g(dz_1^2 + \cdots + dz_4^2),\nn
&& H_{uij} = \mu_{ij}.
\label{d33}
\ena
The corresponding eq.~(\ref{ka4}) for $K$ then becomes
\bea
&&[\Box^{(4)} + g^{-1} \partial_r g \partial_r + H_3 f^{(\c-\b)}g (\pa_{y_2}^2 +\pa^2_{y_3})
\nn
&&\hs{15} + H_{3'} f^{(\c-\b)}(\pa^2_{y_4} + \pa^2_{y_5} )]K
 = - \frac{(\mu_{ij})^2}{2} f^{2\a +2\c +\d} g.
\label{kand33}
\ena
The parameters $\a, \b $ and $\c$ satisfy the relation
\bea
\a+2\b+\c = 0 .
\label{condn33}
\ena
The relation (\ref{cond2n}) becomes
\bea
6\a^2 + 6\c^2 +4\a\c + \d^2 = 3.
\label{cond33n}
\ena
Using these relations and choosing same value for $\a_{\rm D3}$ for both
D3-branes, we find from \p{d33}
\bea
a=\frac{|\a_{\rm D3}|-\a_{\rm D3}}{4}+\a, \quad
b=-\frac{|\a_{\rm D3}|-\a_{\rm D3}}{4}+\c+\frac{1}{2}, \quad
\a_{\rm D3} = 2(\a-\c) = \a_{\rm D3'}.
\ena
We again find that the conditions (\ref{condh}) for the existence of the
horizon cannot be satisfied for real values of parameters $\a, \c,$ and
$\d$ obeying the above conditions. So the non-extremal deformation of
the D3-D3 brane solution also does not admit horizon.

On the other hand, we have examined the Riemann tensors in the parallel
transported frame and found that they again diverge. Thus, both the extremal
and non-extremal solutions do not
have regular horizon. This is contrary to the above result due to the first
criterion that the extremal solution admits a regular horizon.

\subsection{D3-D5 system}

Next let us consider the extremal D3-D5 system. The metric is given by
\bea
ds^2 &=& H_3^{-\frac{1}{2}} H_5^{-\frac{1}{4}}(-2dudv + K du^2 +dy_2^2)
 + H_3^{-\frac{1}{2}}H_5^{\frac{3}{4}} dy^2_3 \nn
&& \hs{5}+\; H_3^{\frac{1}{2}}H_5^{-\frac{1}{4}}(dy^2_4+dy^2_5 +dy_6^2)
+ H_3^{\frac{1}{2}} H_5^{\frac{3}{4}} (dz_1^2 + dz_2^2 + dz_3^2),\nn
&& H_{uij} = \mu_{ij}.
\ena
The corresponding eq.~(\ref{ka4}) for $K$ then becomes
\bea
[\Box^{(3)}  + H_3 H_5 \pa_{y_2}^2 + H_3 \pa^2_{y_3}
+ H_5(\pa^2_{y_4} + \pa^2_{y_5} + \pa^2_{y_6})]K= - \shalf(\mu_{ij})^2.
\label{kaed35}
\ena
where $\Box^{(3)}$ is once again the Laplacian in the three-dimensional
transverse space spanned by $z_1, z_2, z_3$ and $H_3$ and $H_5$ are given by
eq.~(\ref{hae}) with $\tilde d = 1$.

Equation~(\ref{kaed35}) admits a solution of the form
\bea
\hs{-3}
K = c + \frac{\mathcal Q}{r} -\frac{1}{32}(\mu_{ij})^2\Big(r^2 -2(Q_3+2Q_5) r
- 2Q_3 Q_5\ln r + y_2^2+y_3^2 +y_4^2+y_5^2 +y^2_6 \Big).
\ena
So for the extremal D3-D5 solution $a=\frac{3}{8}, b=-\frac{5}{8}$ and
$h=-\shalf$ and the criterion for the existence is not satisfied.

Now let us consider the non-extremal deformation of D3-D5 system. The
metric is given by
\bea
ds^2 &=& H_3^{-\frac{1}{2}} H_5^{-\frac{1}{4}}f^{2\a} (-2dudv + K du^2)
+ H_3^{-\frac{1}{2}} H_5^{-\frac{1}{4}} f^{2\b} dy_2^2
+ H_3^{-\frac{1}{2}}H_5^{\frac{3}{4}} f^{2\b} dy^2_3 \nn
&& \hs{5}+\; H_3^{\frac{1}{2}}H_5^{-\frac{1}{4}}f^{2\b} (dy^2_4+dy^2_5 +dy_6^2)
+ H_3^{\frac{1}{2}} H_5^{\frac{3}{4}} f^{2\c} g^2 (dz_1^2 + dz_2^2 + dz_3^2),
\nn
&& H_{uij} = \mu_{ij}, \quad
e^{\phi}= \tilde H^{-\shalf}_5  f^{\d},
\label{d35}
\ena
where $\tilde H_3$ and $\tilde H_5$ are given by eq.~(\ref{han})
with $\tilde d = 1$.
The corresponding eq.~(\ref{ka4}) for $K$ then gives
\bea
&&[\Box^{(3)} + g^{-1} \partial_r g \partial_r + \tilde H_3 \tilde H_5 f^{2(\c-\b)}g^2\pa^2_{y_2}
+\tilde H_3 f^{2(\c-\b)}g^2\pa^2_{y_3}\nn
&& \hs{15}
+ \tilde H_5 f^{2(\c-\b)}g^2(\pa^2_{y_4} + \pa^2_{y_5}+\pa^2_{y_6} )]K = - \frac{(\mu_{ij})^2}{2} f^{2\a+2\c+\d} g^2.
\label{kand35}
\ena
The parameters $\a, \b, \c$ satisfy the relation (\ref{condnn})
\bea
2\a+5\b+\c = 0 .
\label{cond35}
\ena
The relation (\ref{cond2n}) becomes
\bea
28\a^2 + 8\a\c + 12\c^2 + 5\d^2 = 20.
\label{condn35}
\ena
Using these relations, we find from \p{d35}
\bea
&&a=\frac{|\a_{\rm D3}|-\a_{\rm D3}}{8}+\frac{|\a_{\rm D5}|
-\a_{\rm D5}}{16}+\a, \nn
&&b=-\frac{|\a_{\rm D3}|-\a_{\rm D3}}{8}-\frac{3(|\a_{\rm D5}|
-\a_{\rm D5})}{16}+\c+1, \nn
&&\a_{\rm D3} = \frac{12}{5}\a - \frac{4}{5}\c, \quad
\a_{\rm D5} = \frac{4}{5}\a - \frac{8}{5}\c + \d
\ena
We find that again that the conditions (\ref{condh}) for the existence of
the horizon cannot be satisfied for real values of parameters $\a, \c,$
and $\d$ obeying the above conditions. So the non-extremal deformation of
the D3-D5 brane solution also does not admit horizon.

One also finds that the Riemann tensor in the parallel transported frame for
this solution diverges for both the extremal and non-extremal solutions,
in agreement with the above conclusion.

\subsection{D5-NS5-D3 system}

Next let us consider the extremal D5-NS5-D3 system. The metric is given by
\bea
ds^2 &=& H_3^{-\frac{1}{2}} H_5^{-\frac{1}{4}}H_{5'}^{-\frac{1}{4}}
(-2dudv + K du^2) + H_3^{\frac{1}{2}}H_5^{-\frac{1}{4}}H_{5'}^{-\frac{1}{4}}
(dy^2_2+dy^2_3 +dy_4^2) \nn
&& +\; H_3^{-\frac{1}{2}}H_5^{-\frac{1}{4}}H_{5'}^{\frac{3}{4}}dy^2_5
+ H_3^{-\frac{1}{2}}H_5^{\frac{3}{4}}H_{5'}^{-\frac{1}{4}}dy^2_6
+ H_3^{\frac{1}{2}} H_5^{\frac{3}{4}}H_{5'}^{\frac{3}{4}}
(dz_1^2 + dz_2^2 + dz_3^2),\nn
&& H_{uij} = \mu_{ij}.
\ena
The corresponding eq.~(\ref{ka4}) for $K$ then becomes
\bea
[\Box^{(3)} + H_5 H_{5'} (\pa^2_{y_2}+ \pa^2_{y_3} +\pa_{y_4}^2)
+ H_3 H_5 \pa_{y_5}^2 + H_3 H_{5'}\pa^2_{y_6}]K= - \frac{(\mu_{ij})^2}{2}
H_{5'},
\label{kaed53}
\ena
where $\Box^{(3)}$ is once again the Laplacian in the three-dimensional
transverse space spanned by $z_1, z_2, z_3$ and $H_3,~H_5$ and $H_5'$ are
given by eq.~(\ref{hae}) with $\tilde d = 1$.

Equation~(\ref{kaed53}) admits a solution of the form
\bea
K &=& c + \frac{\mathcal Q}{r} -\frac{1}{32}(\mu_{ij})^2\Big(r^2
-2(2Q_5-2Q_{5'} +Q_{3}) r -6 Q_5 Q_{5'}\ln r - 2 Q_3 Q_5 \ln r \nn
&& \hs{5} -\; 2 Q_3 Q_{5'} \ln r  + y_2^2+y_3^2 +y_4^2+y_5^2 + y_6^2\Big).
\ena
So for the extremal D5-NS5-D3 solution $a=\shalf,b=-1$ and
$h=-\shalf$ and the first criterion for the existence is satisfied.
However, we shall see that this is not sufficient for the existence
of the regular horizon.

Now let us consider the non-extremal deformation of D5-NS5-D3 system. The
metric is given by
\bea
ds^2 &=& H_3^{-\frac{1}{2}} H_5^{-\frac{1}{4}}H_{5'}^{-\frac{1}{4}}
f^{2\a} (-2dudv + K du^2) + H_3^{\frac{1}{2}}H_5^{-\frac{1}{4}}
H_{5'}^{-\frac{1}{4}}f^{2\b} (dy^2_2+dy^2_3 +dy_4^2) \nn
&& +\; H_3^{-\frac{1}{2}}H_5^{-\frac{1}{4}}H_{5'}^{\frac{3}{4}}f^{2\b}
dy^2_5 + H_3^{-\frac{1}{2}}H_5^{\frac{3}{4}}H_{5'}^{-\frac{1}{4}}f^{2\b} dy^2_6
+ H_3^{\frac{1}{2}} H_5^{\frac{3}{4}}H_{5'}^{\frac{3}{4}}
f^{2\c} g^2(dz_1^2 + dz_2^2 + dz_3^2),\nn
&& H_{uij} = \mu_{ij}, \quad
e^{\phi}= \tilde H^{-\shalf}_5 H^{\shalf}_{5'} f^{\d}.
\label{d553}
\ena
where $\tilde H_3, \tilde H_5$ and $\tilde H_{5'}$ are given by
eq.~(\ref{han}) with $\tilde d = 1$.
The corresponding eq.~(\ref{ka4}) for $K$ then gives
\bea
&&[\Box^{(3)} + g^{-1} \partial_r g \partial_r +\tilde H_5\tilde H_{5'} f^{2(\c-\b)}g^2(\pa^2_{y_2}+ \pa^2_{y_3} +\pa^2_{y_4} )+\tilde H_3 \tilde H_5f^{2(\c-\b)}g^2
\pa^2_{y_5} \nn
&& \hs{15}
+\tilde H_3 \tilde H_{5'}f^{2(\c-\b)}g^2\pa^2_{y_6}]K = - \frac{(\mu_{ij})^2}{2}f^{2\a+2\c+\d} g^2 \tilde H_{5'}.
\ena

The parameters $\a, \b, \c,$ satisfy the relation (\ref{condnn}):
\bea
2\a+5\b+\c = 0 .
\ena
The relation (\ref{cond2n}) becomes
\bea
28\a^2 + 8\a\c + 12\c^2 + 5\d^2 = 20.
\label{condn53}
\ena
Using these relations, we find from \p{d553}
\bea
&&a=\frac{|\a_{\rm D3}|-\a_{\rm D3}}{8}+\frac{|\a_{\rm D5}|
-\a_{\rm D5}}{16}++\frac{|\a_{\rm NS5'}|-\a_{\rm NS5'}}{16}+\a, \nn
&&b=-\frac{|\a_{\rm D3}|-\a_{\rm D3}}{8}-\frac{3(|\a_{\rm D5}|
-\a_{\rm D5})}{16}-\frac{3(|\a_{\rm NS5'}|-\a_{\rm NS5'})}{16}+\c+1, \nn
&&\a_{\rm D3} = \frac{12}{5}\a - \frac{4}{5}\c, \quad
\a_{\rm D5} = \frac{4}{5}\a - \frac{8}{5}\c + \d, \quad
\a_{\rm NS5'} = \frac{4}{5}\a - \frac{8}{5}\c - \d
\ena
We find that the conditions (\ref{condh}) for the existence of the horizon
cannot be satisfied for real values of parameters $\a, \c,$ and $\d$ obeying
the above conditions. So the non-extremal deformation of the D5-NS5-D3
system does not admit horizon.

We have examined the Riemann tensors as measured in the parallel transported
frame for this solution and find that they again diverge for both the
extremal and non-extremal solutions.
Thus the above result according to the first criterion is not sufficient to
guarantee the existence of the regular horizon for the extremal solution.

\section{Summary and Discussion}

In this paper we have constructed a general class of intersecting
brane solutions in pp-wave spacetime with nonconstant NS-NS flux.
The intersection rules for the branes in this background are
derived by solving Einstein equations in $D$ dimensional space-time.
Though the asymptotic form of
the solutions is quite different from flat space, the intersection
rules are found to be the same as that of flat space-time.
The supersymmetric properties of the brane configuration are also outlined.
Moreover, we generalize our construction to non-extremal solutions
as well. The corresponding set of intersection rules are also derived.

We have considered the possibility of the existence
of horizon with some examples of intersecting branes in the pp-wave
background with three form flux. Here we have first used the criteria
developed in ref.~\cite{liu}, which indicates that there can be a regular
horizon for some extremal solutions, but the non-extremal deformations do
not admit any horizon. However, the criterion is not sufficient to
guarantee the existence of the horizon~\cite{bre}. We have explicitly
shown that the Riemann tensors diverge in the parallel propagated frame
for the brane solutions in pp-wave background for which the first criterion
suggests that there can exist a regular horizon.

One would conclude that the brane solutions do not admit the regular
horizon in pp-wave spacetime in general and the corresponding solutions
are singular. However, as emphasized in \cite{bre}, the presence of these
'pp-singularities' close off the spacetime near the horizon and analytic
extension beyond the horizon is not possible. In this sense these
singularities may be regarded as the physical boundary of the spacetime.
It is also possible that stringy effects can resolve these singularities.

D-brane systems have been quite useful in studying nonperturbative properties
of field theories realized on them. The pp-wave background is one of the
rare examples on which string theories can be solved exactly. It would be
interesting to examine our solutions further and see how these nice
properties expected to be reflected in our brane solutions can give further
insight into the field theories as well as string theories.

\section*{Acknowledgements}

The work of NO was supported in part by Grants-in-Aid for Scientific Research
Nos. 12640270 and 02041. We would like to thank Dr. Hua Bai for pointing 
out a missing term in the equations for K for nonextreme case, but this 
does not affect any of our conclusions.

\appendix

\section{Curvature tensors}

In this appendix, we summarize the curvature tensors necessary for our
derivation of brane solutions.

The non-zero Christoffel symbols for the metric~\p{met} are
\bea
&&\Gamma^{u}_{ur}=\Gamma^{v}_{vr}= u'_0, ~~~
\Gamma^{v}_{ur} = -\shalf K', ~~~
\Gamma^{r}_{uv} = u'_0 e^{2 u_0 -2 B} ,~~~
\Gamma^{r}_{uu} = -K(u'_0 +\shalf \frac{K'}{K}) e^{2 u_0 -2 B}, \nn
&& \Gamma^{r}_{rr} = B',~~~
\Gamma^{\a}_{\b r} = u'_\a \d_{\a\b} ,~~~
\Gamma^{r}_{\a\b} = - u'_\a e^{2 u_\a -2 B}\d_{\a\b}, \nn
&& \Gamma^{a}_{b r} = (B' + \frac{1}{r}) \d_{ab} ,~~~
\Gamma^{r}_{ab} = - r^2 (B' + \frac{1}{r}) g_{ab} ,~~~
\Gamma^a_{bc}= \shalf g^{ad}(g_{bd,c} + g_{cd,b} -g_{bc,d}),
\ena
where $g_{ab}$ is the metric for the sphere of radius $r$ and
the prime denotes a derivative with respect to $r$.

\vs{4}

{\bf Ricci Tensors:}

The non-zero Ricci tensors of the metric are
\vs{-2}
\bea
R_{uv} &=& e^{2(u_0 - B)} \Bigg[ u_0'' + u_0'\Big\{ 2u_0'
 + \sum_{\a=2}^{d-1} u_\a' + \tilde d B' + \frac{\tilde d+1}{r}
 \Big\}\Bigg], \nn
R_{uu} &=& -e^{2(u_0 - B)}K \Bigg[ u''_0 + \frac{\tilde d+1}{r}u'_0
+ \shalf K^{-1}\Box^{(\tilde d + 2)} K \nn
&& \hs {15}  +\pa_i \Big(u_0+\shalf \ln K\Big) \pa^i\Big(2 u_0
+ \sum_{\a=2}^{d-1} u_\a+ \tilde d B \Big)\Bigg]
- \shalf \sum_{\a=2}^{d-1} e^{2(u_0-u_\a)}\pa^2_\a K , \nn
R_{\a\b} &=& -e^{2(u_\a - B)} \Bigg[ u_\a'' + u_\a'\Big( 2u_0'
+ \sum_{\c=2}^{d-1} u_\c' + \tilde d B' + \frac{\tilde d+1}{r} \Big)
 \Bigg] \d_{\a\b}, \nn
R_{rr} &=& -2u_0'' - \sum_{\a=2}^{d-1} u_\a'' - (\tilde d+1)\Big(B''
- \frac{1}{r^2}\Big) +B'\Big\{ 2u_0' + \sum_{\a=2}^{d-1} u_\a'
+ (\tilde d+1)\Big(B' + \frac{1}{r}\Big)\Big\} \nn
&& -2(u_0')^2 -\sum_{\a=2}^{d-1} (u_\a')^2 -(\tilde d+1)\Big(B'
+ \frac{1}{r}\Big)^2, \nn
R_{ab} &=& - \Bigg[ B'' -\frac{1}{r^2}+\Big(B' + \frac{1}{r}\Big)
\Big( 2u_0' + \sum_{\a=2}^{d-1} u_\a' + \tilde d B' + \frac{\tilde
  d+1}{r} \Big)\Bigg] g_{ab} + \frac{\tilde d}{r^2} g_{ab}.
\label{ricci}
\ena

{\bf Riemann Tensors:}

The nonzero Riemann tensors of the metric are
\vs{-2}
\bea
{R_{uvu}}^v &=& -K {(u'_0)}^2 e^{2u_0 - 2B},\nn
{R_{uvu}}^u &=& - {(u'_0)}^2 e^{2u_0 - 2B},\nn
{R_{uru}}^r &=& \Big(-K u''_0 - K' u'_0 - {K''\over 2} - (u'_0)^2 K
+ K u'_0 B' + {1\over 2} K' B'\Big) e^{2u_0 - 2B}, \nn
{R_{urv}}^r &=& [u''_0 + {(u'_0)}^2 - B' u'_0] e^{2u_0 - 2B},\nn
{R_{u \a u}}^{\b} &=& -K u'_{\alpha} \Big(u'_0 + {1\over 2}
{K'\over K}\Big) e^{2u_0 - 2B}\delta^{\beta}_{\alpha},\nn
{R_{u \a v}}^{\b} &=& u'_{\alpha} u'_{0} e^{2u_0 - 2B}
\delta^{\beta}_{\alpha}, \nn
{R_{u a u}}^{b} &=& -K\Big(B' + {1\over r}\Big)\Big(u'_0 + {1\over 2}
{K'\over K}\Big) e^{2u_0 - 2B} \d_a^b, \nn
{R_{\a r \b}}^{r} &=& -[u''_{\a} +(p +1){u'_{\a}}^2 -
u'_{\a}B'] e^{2u_{\a} - 2B}\d_{\a \b}, \nn
{R_{a r b}}^{r} &=& [ - r^2 B'' + {B'}^2 \tilde{d} r^2
+(2\tilde{d}-1)B' r + \tilde d ] \delta_{ab}, \nn
{R_{a b c}}^{d} &=& \hat{R}_{abc}{}^d -r^2 \Big(B' +{1\over r}\Big)^2
(\delta^d_b \delta_{ac} -
\delta^d_a \delta_{bc}), \nn
{R_{\alpha a \beta}}^{b} &=& -u'_{\alpha} \Big(B' +{1\over r}\Big)
e^{2u_{\alpha} - 2B}\delta^b_a \delta_{\alpha \beta},
\ena
\vs{-2}
where $\hat{R}_{abc}{}^d$ is the Riemann tensor for the sphere part with
metric $g_{ab}$.

\newcommand{\NP}[1]{Nucl.\ Phys.\ B\ {\bf #1}}
\newcommand{\PL}[1]{Phys.\ Lett.\ B\ {\bf #1}}
\newcommand{\CQG}[1]{Class.\ Quant.\ Grav.\ {\bf #1}}
\newcommand{\CMP}[1]{Comm.\ Math.\ Phys.\ {\bf #1}}
\newcommand{\IJMP}[1]{Int.\ Jour.\ Mod.\ Phys.\ {\bf #1}}
\newcommand{\JHEP}[1]{JHEP\ {\bf #1}}
\newcommand{\PR}[1]{Phys.\ Rev.\ D\ {\bf #1}}
\newcommand{\PRL}[1]{Phys.\ Rev.\ Lett.\ {\bf #1}}
\newcommand{\PRE}[1]{Phys.\ Rep.\ {\bf #1}}
\newcommand{\PTP}[1]{Prog.\ Theor.\ Phys.\ {\bf #1}}
\newcommand{\PTPS}[1]{Prog.\ Theor.\ Phys.\ Suppl.\ {\bf #1}}
\newcommand{\MPL}[1]{Mod.\ Phys.\ Lett.\ {\bf #1}}
\newcommand{\JP}[1]{Jour.\ Phys.\ {\bf #1}}


\begin{thebibliography}{99}
\bibitem{klim} D. Amati and C. Klimcik, {\it ``Nonperturbative
 computation of the Weyl anomaly for a class of nontrivial
 backgrounds''}, \PL{219} (1989) 443.
\bibitem{steif}G. T. Horowitz and A. R. Steif, {\it ``Space-time
 singularities in string theory''}, \PRL{64} (1990) 260.
\bibitem{met0} R. R. Metsaev, {\it ``Light cone gauge formulation of
IIB supergravity in $AdS_5 \times S^5$ background and AdS/CFT
 correspondence''}, \PL{468} (1999) 65, hep-th/9908114.
%%CITATION = HEP-TH 9908114;%%
\bibitem{met1}R. R. Metsaev, {\it ``Type IIB Green-Schwarz superstring
 in plane wave Ramond-Ramond background''}, \NP{625} (2002) 70, hep-th/0112044.
%%CITATION = HEP-TH 0112044;%%
\bibitem{met2} R. R. Metsaev and A. A. Tseytlin, {\it ``Exactly solvable
 model of superstring in Ramond-Ramond plane wave background''},
 \PR{D65} (2002) 126004, hep-th/0202109.
%%CITATION = HEP-TH 0202109;%%
\bibitem{penrose} R. Penrose, {\it ``Any space-time has a plane wave as a
 limit''}, in Differential geometry and relativity, pp.271-275,
 Reidel, Dordrecht, (1976).
\bibitem{bfhp} M. Blau, J. Figuero-O'Farrill, C. Hull and
G. Papadopoulos, {\it ``A new maximally supersymmetric background
of IIB superstring theory''}, \JHEP{0201}, 047 (2000), hep-th/0110242,
%%CITATION = HEP-TH 0110242;%%
\bibitem{blau}M. Blau, J. Figuero-O'Farrill and G. Papadopoulos,
{\it ``Penrose limits and maximal supersymmetry''},
\CQG{19} (2000) L87, hep-th/0201081;
%%CITATION = HEP-TH 0201081 ;%%
M. Blau, J. Figuero-O'Farrill and G. Papadopoulos,  {\it ``Penrose limits,
supergravity and brane dynamics''}, \CQG{19} (2002) 4753,
hep-th/0202111.
%%CITATION = HEP-TH 0202111 ;%%
\bibitem{maldacena} J. M. Maldacena, {\it ``The large N limit of
superconformal field theories and supergravity''},
Adv. Theor. Math. Phys. {\bf 2} (1998) 231; Int. J. Theor. Phys.
{\bf 38} (1999) 1113, hep-th/9711200.
%%CITATION = HEP-TH 9711200;%%
\bibitem{malda} D. Berenstein, J. Maldacena and H. Nastase,  {\it ``Strings
in flat space and pp waves from $N=4$ super Yang-Mills''}, \JHEP{0204}
(2002) 013, hep-th/0202021.
%%CITATION = HEP-TH/0202021;%%
\bibitem{li} M. Li, {\it ``PP-wave black holes and the matrix model''},
\JHEP{0305} (2003) 031, hep-th/0212345.
%%CITATION = HEP-TH 0212345;%%
\bibitem{dabh} A. Dabholkar and S. Parvizi, {\it ``Dp Branes in pp-wave
background''}, \NP{641} (2002) 223, hep-th/0203231.
%%CITATION = HEP-TH 0203231;%%
\bibitem{kumar} A. Kumar, R. R. Nayak and Sanjay, {\it ``D-brane
solutions in pp-wave background''}, \PL{541} (2002) 183, hep-th/0204025.
%%CITATION = HEP-TH/0204025;%%
\bibitem{sken} K. Skenderis and M. Taylor, {\it ``Branes in AdS and
pp-wave spacetimes''}, \JHEP{0206} (2002) 025, hep-th/0204054.
%%CITATION = HEP-TH 0204054;%%
\bibitem{bain} P. Bain, P. Meessen and M. Zamaklar, {\it
``Supergravity solutions for D-branes in Hpp-wave backgrounds''},
\CQG{20} (2003) 913, hep-th/0205106.
%%CITATION = HEP-TH 0205106;%%
\bibitem{alis} M. Alishahiha and A. Kumar, {\it ``D-brane solutions
from new isometries of pp-waves''}, \PL{542} (2002) 130, hep-th/0205134.
%%CITATION = HEP-TH 0205134;%%
\bibitem{michi} Y. Michishita, {\it ``D-branes in NSNS and RR pp-wave
 backgrounds and S-duality''}, \JHEP{0210} (2002) 048, hep-th/0206131.
%%CITATION = HEP-TH 0206131;%%
\bibitem{bpz} P. Bain, K. Peeters and M. Zamaklar,
{\it ``D-branes in a plane wave from covariant open strings''},
\PR {67} (2003) 066001, hep-th/0208038.
%%CITATION = HEP-TH/0208038;%%
\bibitem{kamal} A. Biswas. A. Kumar and K. L. Panigrahi,
{\it ``p-p' branes in pp-wave background''}, \PR{66} (2002) 126002,
hep-th/0208042.
%%CITATION = HEP-TH/0208042;%%
\bibitem{stelle} M. Cvetic, H. Lu, C. N. Pope and K. S. Stelle,
{\it ``Linearly-realised worldsheet supersymmetry in pp-wave background''},
hep-th/0209193.
%%CITATION = HEP-TH/0209193;%%
\bibitem{rashmi} R. R. Nayak, {\it ``D-branes at angle in pp-wave
 background''}, \PR{67} (2003) 086006, hep-th/0210230.
%%CITATION = HEP-TH/0210230;%%
\bibitem{alday} L. F. Alday and M. Cirafici, {\it ``An example of localized
D-branes solution on pp-wave backgrounds''}, \JHEP{05} (2003) 006,
hep-th/0301253.
%%CITATION = HEP-TH/0301253;%%
\bibitem{tay} K. Skenderis and M. Taylor, {\it ``Open strings in the
 plane wave background I: Quantization and symmetries''},
hep-th/0211011.
%%CITATION = HEP-TH 0211011;%%
K. Skenderis and M. Taylor, {\it `` Open strings in the plane wave
 background II: Superalgebras and spectra''}, hep-th/0212184.
%%CITATION = HEP-TH 0212184;%%
\bibitem{maoz} J. Maldacena and L. Maoz, {\it ``Strings on pp-waves
 and massive two dimensional field theories''}, \JHEP{0212} (2002) 046,
 hep-th/0207284.
%%CITATION = HEP-TH/0207284;%%
\bibitem{russo} J. G. Russo and A. A. Tseytlin, {\it ``A class of
 exact pp-wave string models with interacting light-cone gauge
 actions''}, \JHEP{0209} (2002) 035, hep-th/0208114.
%%CITATION = HEP-TH/0208114;%%
\bibitem{kim} N. Kim, {\it ``Remarks on type IIB pp waves with
Ramond-Ramond fluxes and massive two dimensional nonlinear
sigma models''}, \PR{67} (2003) 046005, hep-th/0212017.
%%CITATION = HEP-TH/0212017;%%
\bibitem{bonelli}  G. Bonelli, {\it ``On Type II strings in exact
superconformal non-constant RR backgrounds''}, \JHEP{0301} (2003) 065,
 hep-th/0301089.
%%CITATION = HEP-TH/0301089;%%
\bibitem{hikida} Y. Hikida and S. Yamaguchi, {\it ``D-branes in
 pp-waves and massive theories on worldsheet with boundary''},
\JHEP{0301} (2003) 072, hep-th/0210262.
%%CITATION = HEP-TH/0210262;%%
\bibitem{kamal1} K. L. Panigrahi and Sanjay, {\it ``D-branes in pp-wave
spacetime with nonconstant NS-NS flux''}, \PL{561}
(2003) 284, hep-th/0303182.
%%CITATION = HEP-TH/0303182;%%
\bibitem{NO} N. Ohta, {\it ``Intersection rules for non-extreme $p$-branes''},
 \PL{403} (1997) 218, hep-th/9702164.
%%CITATION = HEP-TH/9702164;%%
\bibitem{papado} G. Papadopoulos and P. K. Townsend, {\it ``Intersecting
 M-branes''}, \PL{380} (1996) 273, hep-th/9603087.
%%CITATION = HEP-TH/9603087;%%
\bibitem{tse} A. A. Tseytlin, {\it ``Harmonic superpositions of
 M-branes''}, \NP{475} (1996) 149, hep-th/9604035.
%%CITATION = HEP-TH/9604035;%%
\bibitem{bbj} K. Behrndt, E. Bergshoeff and B. Janssen,
{\it ``Intersecting D-Branes in ten and six dimensions''}, \PR{55}
(1997) 3785, hep-th/9604168.
%%CITATION = HEP-TH/9604168;%%
\bibitem{gkt} J. P. Gauntlett, D. A. Kastor and J. Traschen,
{\it ``Overlapping branes in M theory''}, \NP{478} (1996) 544,
hep-th/9604179.
%%CITATION = HEP-TH/9604179;%%
\bibitem{aeh} R. Argurio, F. Englert and L. Houart, {\it ``Intersection
rules for $p$-branes''}, \PL{398} (1997) 61, hep-th/9704190;
%%CITATION = HEP-TH/9704190;%%
\bibitem{NO1} N. Ohta, {\it ``Intersection rules for S-branes''},
 \PL{558} (2003) 213, hep-th/0301095;
%%CITATION = HEP-TH/0301095;%%
N. Ohta, {\it ``Null-brane solutions in supergravities''},
 \PL{559} (2003) 270, hep-th/0302140.
%%CITATION = HEP-TH/0302140;%%
\bibitem{rangamani} V. E. Hubeny and M. Rangamani,
{\it ``No horizons in pp-waves''}, \JHEP{0211} (2002)
021, hep-th/0210234.
%%CITATION = HEP-TH/0210234;%%
\bibitem{ranga1}V. E. Hubeny and M. Rangamani, {\it ``Causal structures
 of pp-waves''}, \JHEP{0212} (2002) 043, hep-th/0211195.
%%CITATION = HEP-TH/0211195;%%
V. E. Hubeny and M. Rangamani,  {\it ``Generating asymptotically
 plane wave spacetimes''}, \JHEP{0301} (2003) 031, hep-th/0211206.
%%CITATION = HEP-TH/0211206;%%
\bibitem{liu}  J. T. Liu, L. A. Pando Zayas and D. Vaman,
{\it ``On horizons and plane waves''}, hep-th/0301187.
%%CITATION = HEP-TH/0301187;%%
\bibitem{gimon} E. G. Gimon, A. Hashimoto, V. E. Hubeny, O. Lunin and
M. Rangamani, {\it ''Black strings in asymptotically plane wave geometries''},
hep-th/0306131.
%%CITATION = HEP-TH/0306131;%%
\bibitem{bre} G. T. Horowitz and H. Yang, {\it Black strings and classical
 hair}, \PR{55} (1997) 7618, hep-th/9701077;\\
D. Brecher, A. Chamblin and H.S. Reall, {\it ``AdS/CFT in the infinite
 momentum frame''}, \NP{607} (2001) 155, hep-th/0012076.
%%CITATION = HEP-TH/0012076;%%
\bibitem{pope} M. Cvetic, H. Lu and C. N. Pope, {\it ``M-theory pp-waves,
Penrose limits and supernumerary supersymmetries''},
\NP{644} (2002) 65, hep-th/0203229.
%%CITATION = HEP-TH/0203229;%%
\end{thebibliography}
\end{document}